\documentclass[10pt,twocolumn, showpacs, showkeys]{revtex4-1}
\usepackage{epstopdf}
\usepackage{graphicx}
\usepackage[]{natbib}

\begin{document}

\bibliographystyle{plainnat}

\title{High Magnetic Shear Gain in a Liquid Sodium Stable Couette 
Flow Experiment \\
A Prelude to an $\alpha - \Omega$ Dynamo}

\author{Stirling A. Colgate$^1$,$^2$}

\author{Howard Beckley$^2$, (deceased),  Jiahe Si$^2$, Joe Martinic$^2$, David  Westpfahl$^2$,  James Slutz$^2$, Cebastian Westrom$^2$,  Brianna Klein$^2$, Paul Schendel$^2$, Cletus Scharle$^2$,  Travis McKinney$^2$, Rocky Ginanni$^2$, Ian Bentley$^2$,  Timothy Mickey$^2$, Regnar Ferrel$^2$}

\author{Hui Li$^1$ Vladimir Pariev$^1$, John Finn$^3$}

\affiliation{$^1$T-2, MS B-227, Los Alamos National Laboratory, Los Alamos, NM 87545; colgate@lanl.gov}

\affiliation{$^2$Department of Physics, New Mexico Institute of Mining and
Technology, Socorro, NM 87801}

\affiliation{$^3$T-5,  Los Alamos National Laboratory, Los Alamos, NM 87545;} 

\begin{abstract}
The $\Omega$-phase of the liquid sodium $\alpha$-$\Omega$ dynamo experiment  at NMIMT in cooperation with LANL has been successfully demonstrated to produce a high toroidal field $B_{\phi}$ that is $\simeq 8\times  B_r$, where $B_r$  is the radial component of an applied   poloidal magnetic field.  This enhanced toroidal field is produced by the rotational shear  in  stable Couette flow within liquid sodium at magnetic Reynolds number $Rm \simeq 120$.   The small turbulence in stable Taylor-Couette flow is caused by Ekman flow with an estimated turbulence energy fraction of $(\delta v/v)^2 \sim 10^{-3}$.  This high  $\Omega$-gain in low turbulence flow contrasts with a smaller $\Omega$-gain in higher turbulence shear flows. This result supports the ansatz that large scale astrophysical magnetic fields can be created  by semi-coherent large scale motions in which turbulence plays only a smaller diffusive  role that enables magnetic flux linkage. 
\end{abstract}



\maketitle

Major efforts are spent  world-wide on
astrophysical  phenomena that depend upon magnetic fields, e.g., planetary, solar, and stellar magnetic fields, X-rays, cosmic rays, TeV gamma rays, jets and radio lobes powered by 
active galactic nuclei (AGNs), and in the interpretation of Faraday rotation maps of galaxy clusters.  Yet, so far, there is no universally accepted explanation for the inferred larage-scale  
magnetic field. A process, called the  $\alpha-\Omega$  
dynamo \citep{parker55,mof78,krause80}, has been proposed which involves two orthogonal conducting fluid  motions, shear and helicity.  When the two motions are comparable, it is often described 
as a stretch, twist, and fold or "fast" dynamo, and supposedly can be produced by turbulence alone \citep {nor92}, \citep{zeldovich83}.  The problem is that a turbulent dynamo must create these two orthogonal motions turbulent motions.   Fluid turbulence maximizes entropy and the diffusion of the flux of momentum \citep{tob08}, so that a field twisted one way by turbulent eddies at one moment of time may be partially twisted the opposite way  with only  a net bias, and  hence partially diffusive.

One possible way to achieve dynamo is to use naturally occurring large scale near-stable  rotational shear flows (as in AGN accretion disks and in stars) in combination with transverse, transient, rotationally coherent  plumes.  Such plumes may be driven either by star--disk collisions in AGNs \citep{par07a} or large scale (density scale height) convective elements in the base of the convective zone in stars \citep{cha60, wil94, mestel99}.  These flows can stretch and wind-up an embedded, transverse magnetic flux  through a large number of turns (the $\Omega$ effect).  In both types of shear flows the turbulence is expected to be relatively small because of the stability imposed by either an angular momentum gradient or an entropy gradient.  (Positive work must be done to turn over an eddy in such a gradient.)   The  advantage of a large $\Omega$-gain is that the corresponding $\alpha$-gain (the helicity deformation necessary to convert a fraction of the toroidal flux back into poloidal flux) can be correspondingly  smaller.
Parker, the originator of the $\alpha - \Omega$ dynamo concept for astrophysical dynamos,  invoked cyclonic motions to produce the helicity, \citep{parker79}. Moffatt \citep{mof78} countered that a typical atmospheric cyclone makes very many turns before dissipating, therefore making the sign of the resulting poloidal flux incoherent.  Moffat sought motions where the rotation angle was finite, $\sim \pi/4$.  Willette \citep{wil94} and the first author suggested that buoyant plumes were a  unique solution to the problem of coherent dynamo helicity.  Beckley et al. \citep{bec03} demonstrated this property  experimentally of finite, $\sim \pi/4$, coherent rotation of a driven plume before it dissipated through turbulent diffusion in the background fluid.  

Two liquid sodium dynamo experiments have produced
positive exponential gain, but the flows  were constrained by rigid walls unlike astrophysical flows  \citep{stig01,gail00}.  The rigid walls restrict turbulent eddy size by the distance from the wall, log-law of the walls \citep{lan59},
 thereby producing low turbulence. The three recent experiments use the Dudley-James \citep{dud89} or Von Karman flow (counter rotating flow converging  at the mid-plane and  driven by two counter rotating propellers or two vaned turbine impellers, respectively, \citep{for02, norn06, Spen07, lat01, pet03, odi98,  pef00}).  Turbulence is induced by the Helmholtz instability at the shearing mid-plane.  This combination of coherent motions and unconstrained shear-driven (relatively high level) turbulence resulted in a maximum $\Omega$-gain of only
$\sim  \times 2$.  Recognizing the enhanced resistivity of the mid-plane turbulence, the team at Wisconsin added a mid-plane baffle to reduce the turbulence,  following which the $\Omega$-gain increased to $\simeq \times 4$ \citep{rah10}. The von Karman Sodium 2 (VKS2) experiment in the same geometry produced exponential gain \citep{mon07};  However, the dynamo action was explained not by turbulence but primarily by the production of helicity by the large coherent vortices  produced by the radial rigid vanes of the impeller   \citep{lag08}.  (Ferro magnetism added additional gain to this source of helicity.)

The New Mexico $\alpha-\Omega$ dynamo (NMD) experiment  is a
collaboration between the New Mexico Institute of Mining and
Technology and Los Alamos National Laboratory  \citep{col02, col06}.  It
is designed to explore the possibility in the  laboratory using coherent fluid motions in low-turbulence high shear Couette flow  in the annular volume 
between two coaxial cylinders, $R_2/R_1 = 2$  rotating 
at different angular velocities (see Figure 1), closely analogous to natural fluid motions that occur in astrophysical bodies \citep{Bra98, par07a,par07b}. 
The results of the first phase ($\Omega$-phase) of  this two-phase experiment are presented here. 

Fig. 1 also indicates the Ekman flow, a thin fluid layer flowing along the boundaries of the annular volume and including a small radial return flow from inner to outer cylinder through the fluid volume. The Ekman flow produces both a torque and a small but finite level of turbulence \citep{ji01}. This turbulence adds a small turbulent resistivity to the  resistivity of metallic sodium. Five pressure sensors are distributed radially at one end. They measure the pressure distributions, giving information of the actual rotational flow angular velocity distribution. 

\begin{figure}[ht]
\includegraphics[width=2in]{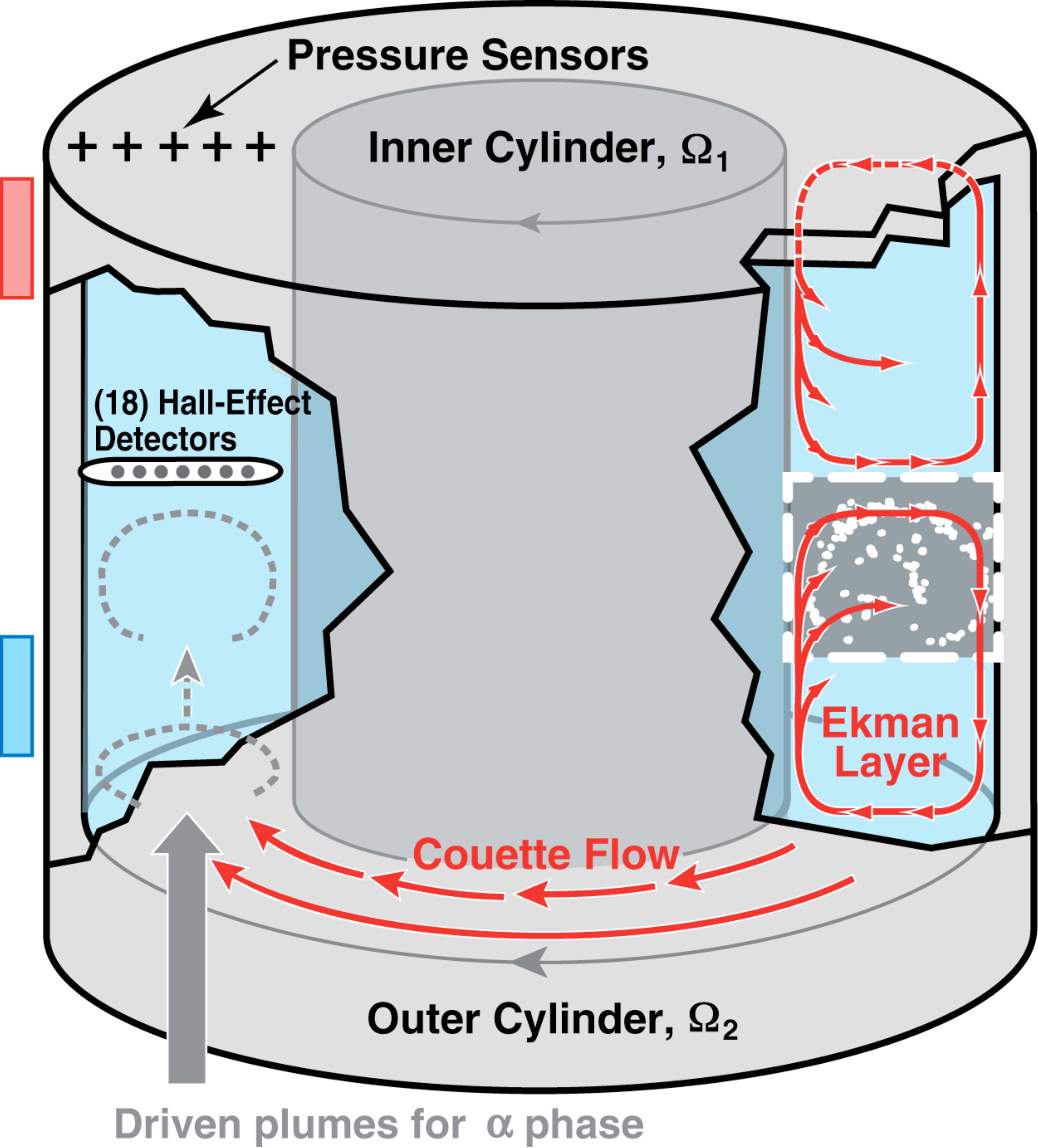}
\caption {A schematic drawing  of the fluid flow of the experiment shows the inner cylinder of $R_1 = 15.2$ cm rotating at $\Omega_1/2\pi = 68$Hz relative to the outer or confining cylinder of  $R_{2} = 30.5$ cm at $\Omega_2 = \Omega_1/4 $.  Liquid sodium in the annular space undergoes Couette rotational flow. In addition, a thin layer in contact with the cylinder walls and end walls undergoes Ekman flow, circulating orthogonally around and then through the Couette flow.  Pressure sensors in the end wall and the magnetic probe housing are also shown schematically.}  
\label{fig1}
\end{figure}

For the magnetic measurements, a 50 kW AC induction motor (20kW is used for stable Couette flow) drives a gear train with clutches and power take-off that rotates the two cylinders at the fixed ratio of $\Omega_1 = 4 \Omega_2$.  The outer cylinder can also be disengaged from the gear train by a clutch, and a DC motor is used to accelerate or brake the outer cylinder independently from the driven inner cylinder.   This allows different $\Omega_1 /\Omega_2$ ratios to be explored.  The DC motor housing (stator) is mounted on bearings. A torque arm with two force sensors ($+-$) connects the motor stator to ground, so that the torque on the DC motor can be measured separately from the drive of the inner cylinder. This arrangement allows us to measure the torque between the two cylinders due to the Ekman flow along the surface of the end plates which rotate more slowly at $\Omega_2$. 

In particular, when the inner cylinder is driven at higher speed by the AC motor,
the Ekman fluid torque tries to spin up or accelerate the outer cylinder.  Two torques counteract this acceleration: 1) the friction in the bearings that support the rotation of the outer cylinder and 2) the torque on the DC motor when used as a generator, or brake.  (The generated power is dissipated in a resistor.)  
In Fig. 2 (left) the crosses are the measurements of the braking force exerted by the DC motor torque arm; the dashed line is the calibrated  bearing torque (measured by disengaging the inner cylinder drive and rotating the outer cylinder with the DC motor alone). The sum of these two torques is equal to the Ekman fluid torque spinning up the outer cylinder.

\begin{figure}[h]
\includegraphics[width=3.5in]{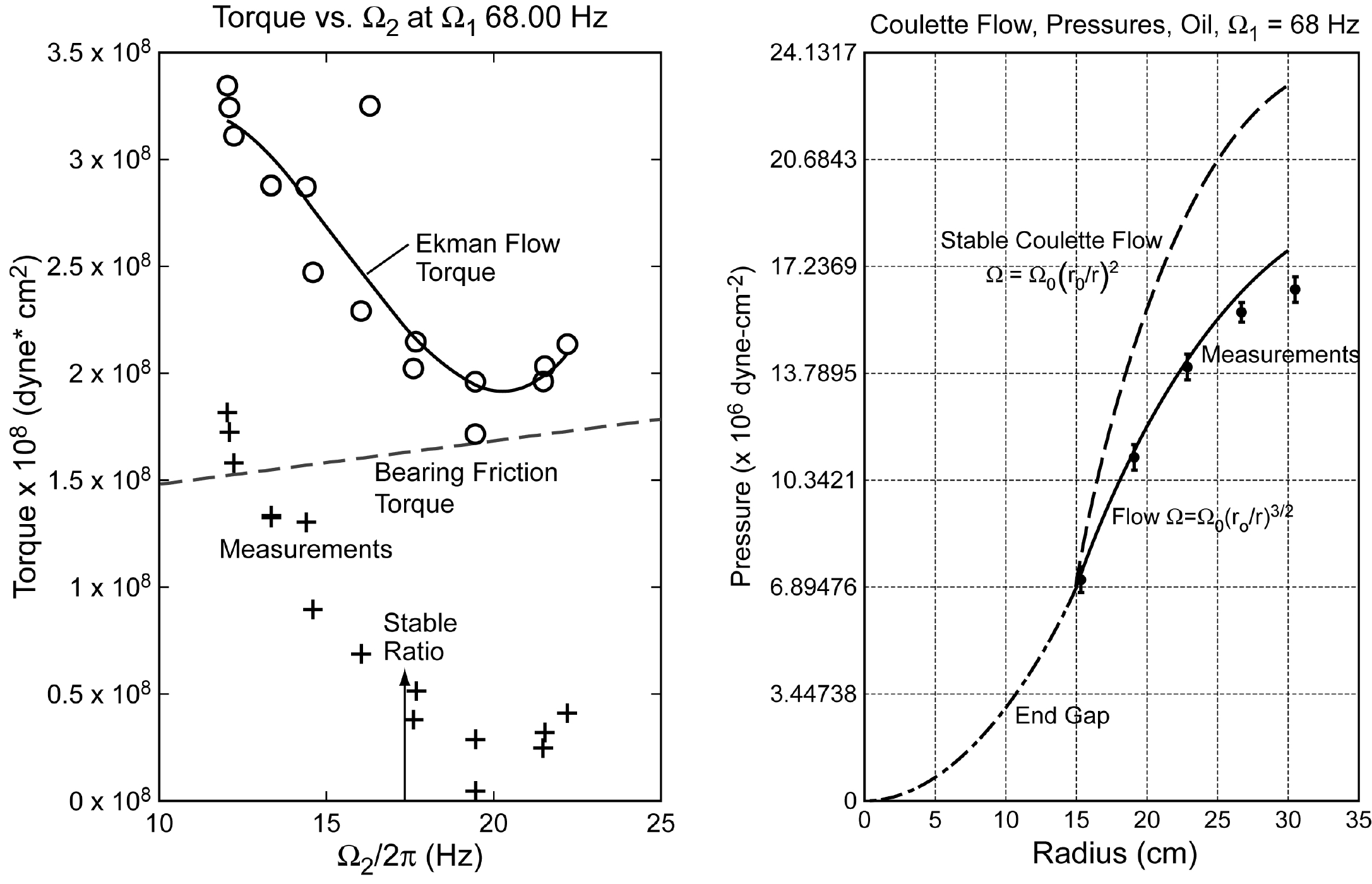}
\caption{(left) The lower  crosses are the measured DC generator torques; the dashed line is the measured bearings torque, and the sum of the two is the top  solid curve, the torque due to the Ekman flow.   
(right) Shown is the measured and theoretical pressure distribution.  The pressure measurement starts at $R_1$  and extends to $R_2$.  
The finite pressure at $R_1$, $P_1 \simeq 100$ psi is due to the 
centrifugal pressure of rotating fluid in the narrow gap between the inner cylinder rotating at $\Omega_1$ and  the end wall rotating at $\Omega_2$  (dot-dash curve).     This thin end-gap is filled with rotating fluid, which extends from the shaft seal (where $P=1$ atm) at radius, $R_3 \simeq R_1/3$ to $R_1$ at which radius the Couette flow in the annular volume is initiated.   From there, the pressure in the annular volume is extrapolated either as $\Omega  \propto R^{-2}$ or   $R^{-3/2}$. The measurements fall along the $R^{-3/2}$ pressure distribution curve indicating less shear than the ideal Couette flow.}
\label{fig2}
\end{figure}
 
The experiment was originally designed upon the supposition that the  primary torque between the two cylinders would be due to the radial Ekman flow at the end walls and would be small, $\sim 1/10$, compared to  pipe flow wall friction stress, $\tau = \rho C_D v^2$.  In  Fig. 2 (left)   the minimal Ekman flow torque occurs  at $\Omega_1/\Omega_2 \sim <4$, somewhat less than the limit of stable Couette flow.   The torque value is $2\times 10^8$ dyne-cm, close to the approximation that each Ekman layer thickness  \citep{ji01}  is  
$\Delta \simeq R_1 /Re^{1/2} \simeq 5 \times 10^{-3}$ cm and  that the average radial flow velocity within the layer is $v_r \simeq (\Omega_1 R_1) /2$. 
In Fig 2 (right),  the measured pressures are compared to the calculated pressures corresponding to two different angular velocity power laws.   The upper (dashed) curve corresponds to ideal, maximum shear, stable Couette flow, or $\Omega \propto r^{-2}$, but  the experimental points follow the lower solid curve corresponding to 
$\Omega \propto r^{-3/2}$. The Ekman flow torque has distorted the angular velocity profile and reduced the shear relative to the ideal Couette flow.

Turbulent flow at such high Reynolds number  $Re \simeq10^7$ is still well beyond current simulation capability.  The Ekman flow is a flux of fluid of reduced angular momentum deposited at the inner cylinder.  The torque  reducing this angular momentum is friction with the end walls, which in turn reduces the angular velocity of the Couette flow in the annular volume.  Unstable flow, turbulent friction with the inner cylinder surface, counteracts this torque by speeding up this flow.   The torque in the Couette volume is a constant, independent of radius and axial position.   Then this shear stress is maintained constant in two laminar sub layers \citep{sch60} with $Re \simeq 100$ and in two turbulent boundary layers in a log-law-of-the-walls solution \citep{lan59}.  When the distance from the walls corresponds to an eddy scale of the radial gradient,  a transition takes place to a scale independent eddy size turbulent torque connecting the two regions at inner and outer boundaries.  
The  mean velocity distribution in the above analysis can give a reduced shear, 
$\Omega / \Omega_1 \propto (R_1/R)^{1.64}$, rather than stable Couette  flow. 
  
 \begin{figure}[ht]
 \includegraphics[width=3in]{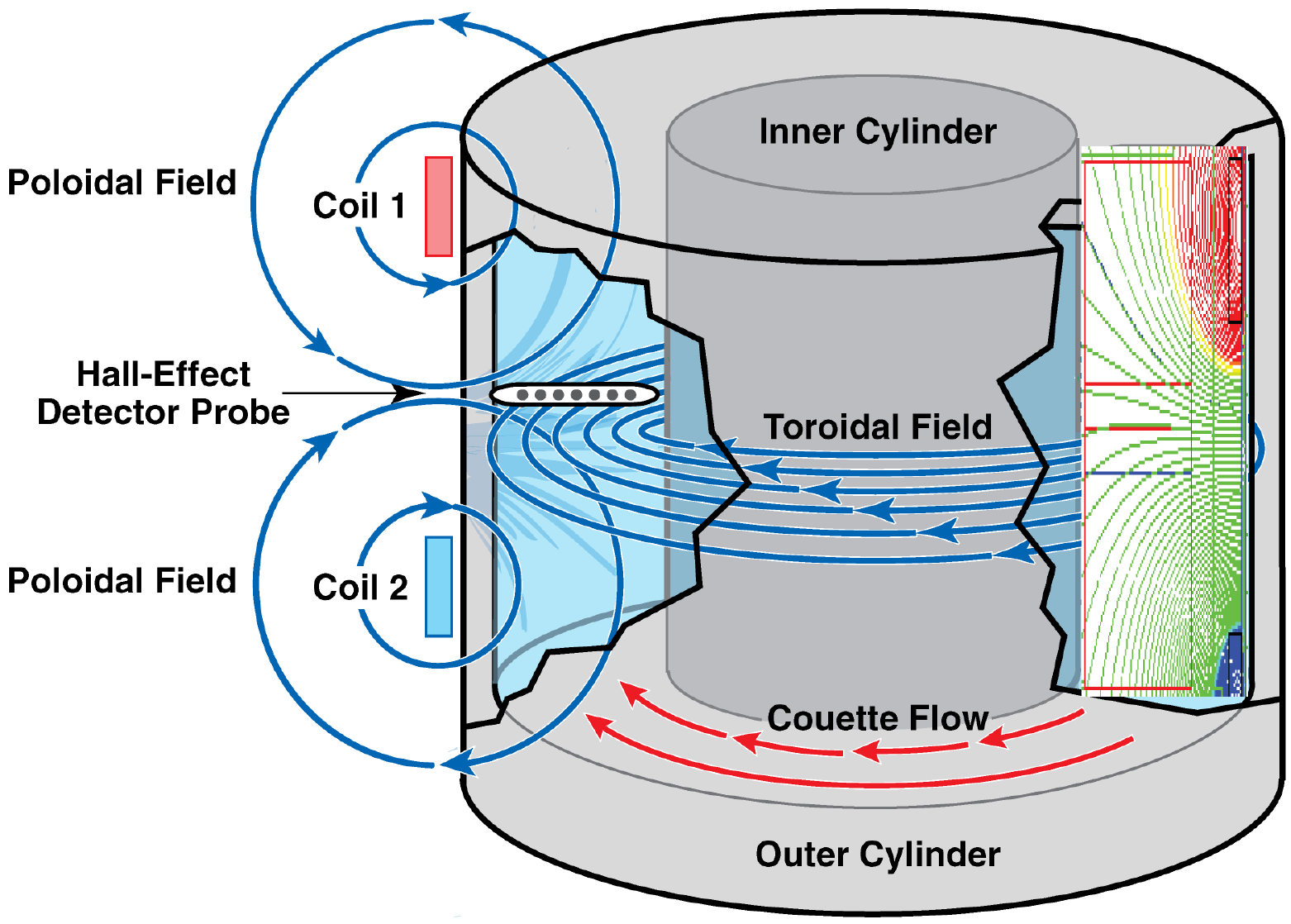}
\caption {A schematic drawing of the poloidal magnetic flux produced by two coils  (left).  Superimposed (right) are the flux lines from the Maxwell calculation \citep{max80},  with the iron shield and steel shaft included.  
This radial flux crosses the high  shear of the Couette flow  producing the enhanced toroidal field.  The $B_r$ field from the Maxwell  calculations agrees with the calibrated  probe measurements to $10\%$. }  
\label{fig3}
\end{figure} 

Fig. 3 shows the magnetic field configuration.  The magnetic field Hall detector probe, internal to the liquid sodium at the mid-plane, is  the primary diagnostic of the experiment.  It consists of 6 multiple, 3-axis, magnetic field Hall effect detectors at the mid-plane in the annular space between the two cylinders and contained in an aero-dynamically shaped housing.  (The fluid friction drag produced by this housing, primarily Ekman flow, is estimated to be $\sim 0.1$ of the end-wall Ekman torque.)   The $\Omega$-gain was then measured using an applied calibrated 
$B_r$ magnetic  field as a function of the coil currents.  Because of the high  gain in $B_{\phi}$ and the lack of perfect orthogonality of the Hall detectors, the measured $B_r$ would be expected to be contaminated by a small fraction of the much larger $B_{\phi}$ field.  

\begin{figure}[h]
 \includegraphics[width=3.5in]{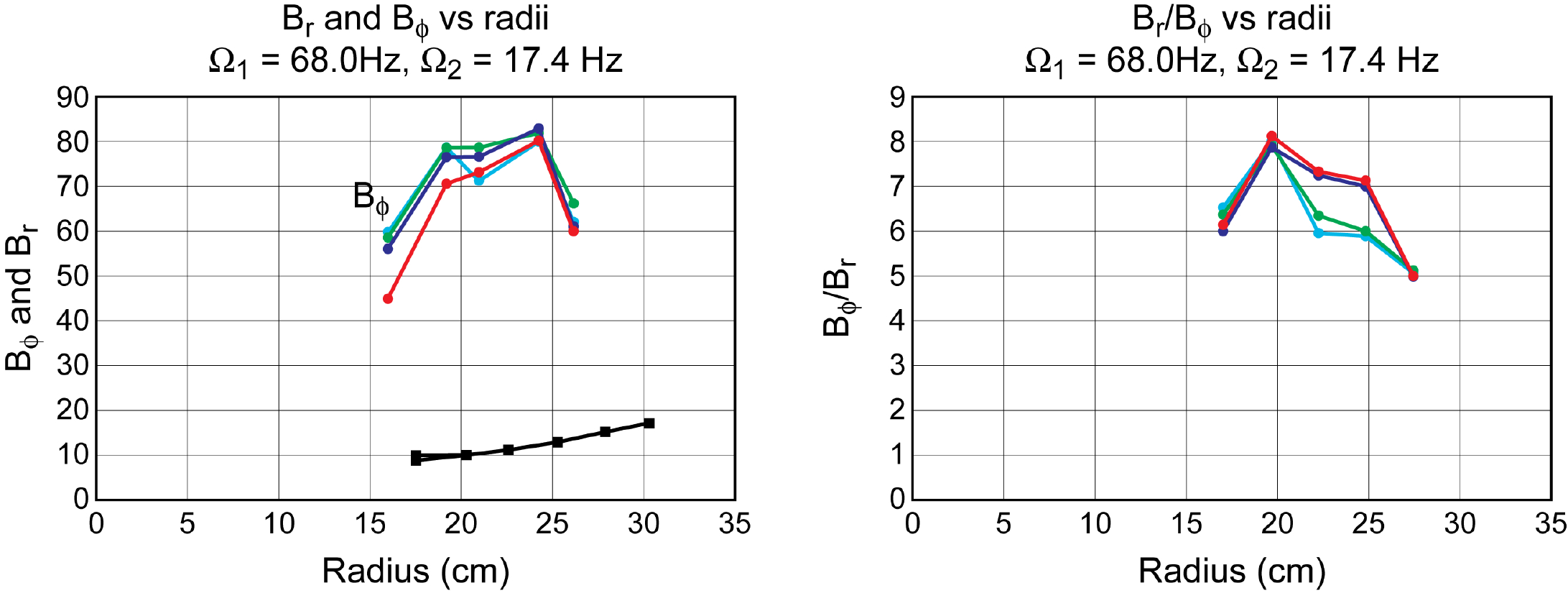}
\caption{ (left) The lower curve shows the applied radial field $B_r \simeq 12$ G and the upper curves show the measured toroidal magnetic field for four over-laid experiments  at  
$\Omega_1 = 68$ Hz. The right  panel  shows the ratio $B_{\phi}/B_r$.}
\label{fig4}
\end{figure}
\begin{figure}[ht]
\includegraphics[width=2in]{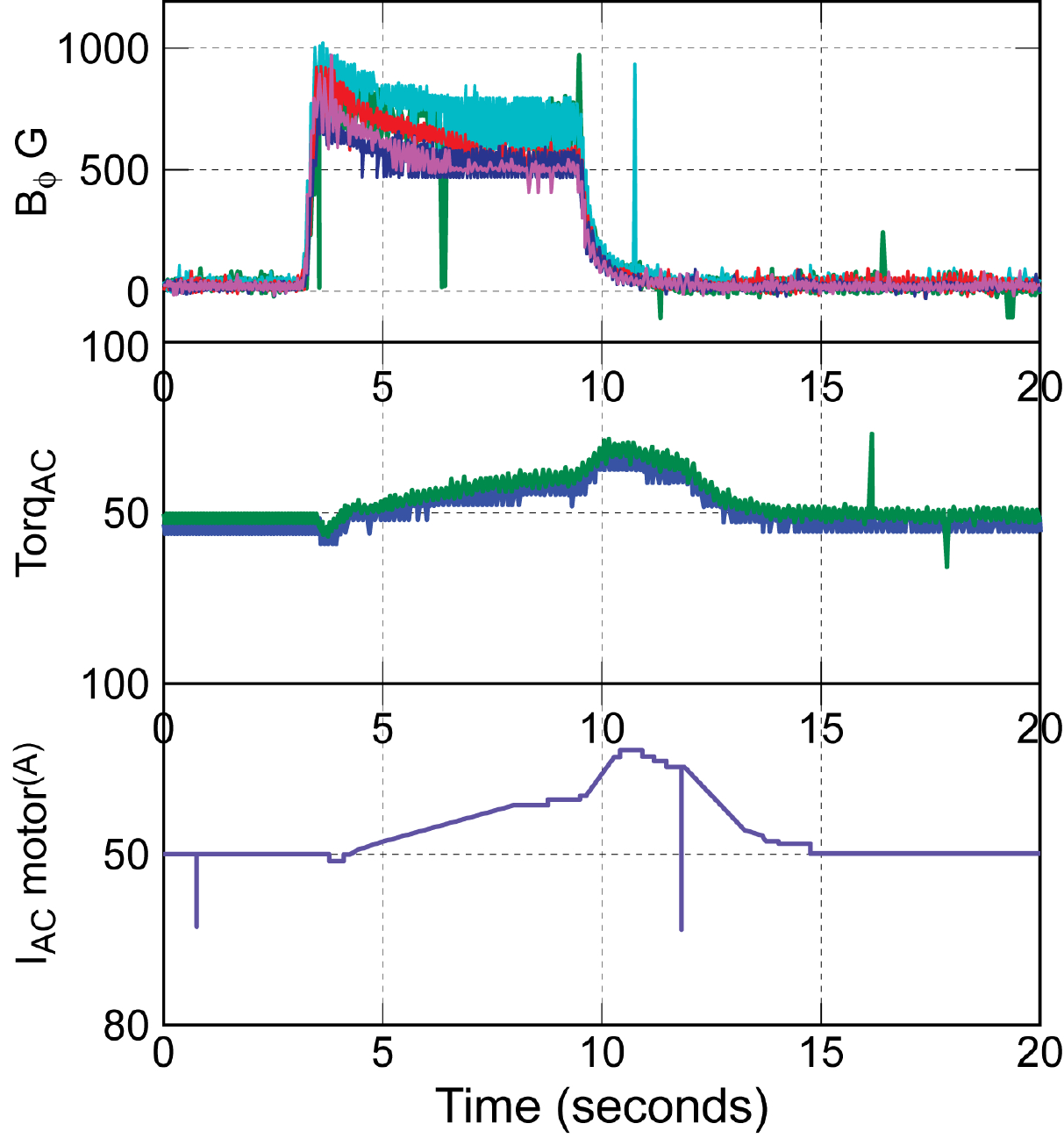}
\caption{The top trace shows the toroidal field $B_{\phi}\sim 750$G vs. time from an applied radial field of $\sim 250$ G.  The $\Omega$-gain is now reduced to $\times 3$.  The bottom traces  are the AC motor torque and current respectively, each showing a $50\%$ increase due to back reaction.  The delay of several seconds in the back reaction torque  corresponds to the spin-down time for the Couette flow to reach a new, modified velocity profile.}
\label{fig5}
\end{figure}

Fig 4 confirms that the measurements are repeatable by showing  four experiments over-laid where the  $\Omega$-gain ratio of $\times 8$ is repeated all  with a low bias field of 
$ B_r\simeq 12$ G.  The  $B_{\phi}$  produced by the Couette  flow shear is about
$\simeq 8 \times B_r$.  Note that the  ideal $\Omega$-gain could be as high as $\sim R_m/4\pi \simeq \times 20$ \citep{par07b}, but 
the reduced shear in the sodium (i.e., not exactly Couette profile) partially  accounts for the reduction in the $\Omega$-gain.
We then measured whether an increase in the velocity shear in the Couette volume might increase the  $\Omega$-gain ratio, as one might naively expect.  The velocity shear is increased by slowing the outer cylinder by using the DC motor as a generator or brake.  Therefore the ratio $\Omega_1/\Omega_2$ was increased by 20\%,  but the  $\Omega$-gain remains constant.  This suggests the gain expected due to shear is balanced by the greater flux diffusion due to the increased turbulence of the now unstable Couette flow (see Fig 2).  
Roughly 100 electronic samples are recorded and averaged during the several seconds recording cycle of each of four magnetic  experiments.  The time variation  between each run reflects slight changes in  the Couette flow relaxation time and hence, the angular velocity distribution.  The repeatability among these four runs as well as several earlier runs in the previous six months gives us confidence that the conclusion of high $\Omega$-gain is valid.

The expected back-reaction should occur at a larger value of $B_r$ where the dissipation and torque of the enhanced 
 $B_{\phi}$ affects the flow field and the added torque shows up  as additional current   and torque in the AC drive motor.  This effect is shown in Fig 5, where, in addition, the expected reduction in the  $\Omega$-gain ratio, from $\times 8$ to $\sim \times 3$ is observed when the applied poloidal field is $B_r \simeq 250$G.  The back reaction stress will  modify the Couette flow profile such as to reduce the  $\Omega$-gain  and hence reduce the back reaction stress until a new steady state is achieved at reduced $B_{\phi}$.  Fig 5 shows the measured difference in power, $\simeq 10$kW, due to the back reaction.  We estimate that the specific magnetic field energy density,  $B_{\phi}^2/8\pi$ is dissipated at  $<\Omega> \times(Vol)$ where the volume of high $B_{\phi}$ is estimated as 1/3 the length, $Vol \simeq (L/3) (\pi/2) (R_2^2 - R_1^2)$.  This results in a power of $\sim 8$ kW, roughly the measured 10 kW.
 
In conclusion, a large  $\Omega$-gain in low turbulent shear in a conducting fluid has been demonstrated. This is likely to be the mechanism of the $\Omega$-gain of a coherent $\alpha-\Omega$ astrophysical dynamo. Driven plumes are planned to be used in the $\alpha$ phase of the NMD experiment.

This experiment has been funded by NSF,  Univ. of Calif. \& LANL, IGPP LANL, and NMIMT.  

\clearpage

\end{document}